\documentclass{article}
\usepackage{spconf,amsmath,graphicx}
\usepackage{graphicx}
\usepackage{booktabs}
\usepackage{siunitx}
\usepackage{cite}
\usepackage{comment}
\usepackage{multirow}
\usepackage[]{graphicx}
\usepackage{subfig}
\usepackage{verbatim}
\title{PosCUDA: Position based Convolution for Unlearnable Audio Datasets}
\twoauthors
 {Vignesh Gokul}
	{University of California, San Diego\\
	Department of  Computer Science\\
    vgokul@ucsd.edu}
 {Shlomo Dubnov}
	{University of California, San Diego\\
	Department of Music\\
    sdubnov@ucsd.edu}

\begin{document}
%\ninept
%
\maketitle
\begin{abstract}
Deep learning models require large amounts of clean data to acheive good performance. To avoid the cost of expensive data acquisition, researchers use the abundant data available on the internet. This raises significant privacy concerns on the potential misuse of personal data for model training without authorisation. Recent works such as CUDA propose solutions to this problem by adding class-wise blurs to make datasets unlearnable, i.e a model can never use the acquired dataset for learning. However these methods often reduce the quality of the data making it useless for practical applications. We introduce PosCUDA, a position based convolution for creating unlearnable audio datasets. PosCUDA uses class-wise convolutions on small patches of audio. The location of the patches are based on a private key for each class, hence the model learns the relations between positional blurs and labels, while failing to generalize. We empirically show that PosCUDA can achieve unlearnability while maintaining the quality of the original audio datasets. Our proposed method is also robust to different audio feature representations such as MFCC, raw audio and different architectures such as transformers, convolutional networks etc.
\end{abstract}
\begin{keywords}
Unlearnable Datasets, Adversarial Learning, Audio Classification
\end{keywords}
\section{Introduction}
\label{sec:intro}

With the advent of large deep learning models, there is a huge need for massive clean datasets. To offset the expensive cost of data collection, researchers use the widely abundant data available on the internet. This includes image, text and audio data that individual users have uploaded to the internet. This raises an important problem on unauthorized usage of private personal data for model training. For instance, recently image generation models have been trained on facial images, artistic content etc that are available on the internet without the consent of the owners \cite{hill2022secretive}. Similarly, musical audio samples of different content creators have been used for training both classification and generative models \cite{drott2021copyright,sturm2019artificial}. All these incidents emphasize the need to create unlearnable datasets that can be uploaded by content creators while being protected from being consumed by learning models.
One approach to solve this issue is by creating unlearnable datasets\cite{fowl2021adversarial,huang2021unlearnable, yuan2021neural}, i.e datasets that tamper any model's capability to learn from them. If an attacker builds a model using unlearnable datasets, the performance of such a model should be similar to a random baseline. There are three important characterstics of unlearnable datasets: 1) creation of unlearnable datasets should be fast and 3) the quality of unlearnable datasets should be as close as possible to the original data for practical purposes. Recent works such as REM \cite{fu2022robust} and CUDA \cite{sadasivan2023cuda} try to learn unlearnable datasets that satisfy all the three criteria. However REM has been shown to be vulnerable in the presence of data augmentation and CUDA affects the quality of the data by a huge factor. Moreover, all the above methods focus on image datasets and are not applicable for audio data.
Audio unlearnable datasets present a unique challenge as the quality of audio can be easily detoriarated with additive noises. Furthermore, machine learning models for audio use a variety of input representations such as MFCC, spectrograms, raw audio etc that can easily break any pre-programmed unlearnability in the dataset. It is important for audio unlearnable datasets to be robust to any feature representation derived from the original dataset and data augmentations, while preserving the quality of the original audio as much as possible. 

To address the above limitations, we propose PosCUDA, a method to create unlearnable datasets for audio. PosCUDA uses positional class-wise filters to help model learn relationships between labels and data, failing to generalize on new data. We apply class-wise blurs to a patch location of a sample based on a private key for each class. This means that instead of applying noise/blur over the entire data, only a small portion of the data is blurred, preserving the overall quality of the original sample. PosCUDA is fast, robust to any feature representation and minimally affects the quality of data making it applicable for practical use. 

\section{Related Work}
\label{sec:relatedwork}

\textbf{Adversarial Poisoning:} Poisoning attacks\cite{chen2017targeted,li2021neural,liu2020reflection,nguyen2020input} introduce some form of noise in the training process to make the model fail or mislabel classes intentionally during testing. Adversarial Poisoning has been widely used to create unlearnable datasets. However, these attacks works only for examples that have noise/trigger patterns in them. This means that during test time, the data has to have the appropriate noise and trigger patterns for adversarial poisoning to work.\\
% \textbf{Adversarial Training:} Adversarial Training is a method to train models that are robust to adversarial poisons. It involves training models with data that includes adversarial examples. \\
\textbf{Data Privacy:} Data privacy methods \cite{dwork2006differential, abadi2016deep} preserve the model from leaking data that has been used for training\cite{mugunthan2021dpd}. Often this includes setups where multiple parties train models together sharing private data. Unlearnable datasets differ from privacy protection, as the task is to make the dataset non-usable by any model.\\
\textbf{Unlearnable Datasets:} There have been significant research in developing methods to create unlearnable datasets. \cite{huang2021unlearnable} introduced an error-minimizing noise that reduces the training error of a class to zero, which prevents the model from learning useful features from those examples. Targeted Adversarial Poisoning (TAP) \cite{fowl2021adversarial} use error-maximizing noises as data poisons to attack the model. Neural Tangent Generalization Attacks \cite{yuan2021neural} generates label attacks to detoriorate the generarlization capability of models trained on such data. Robust Error Minmization (REM) \cite{fu2022robust} is a method to produce unlearnable datasets that are robust to adversarial training. REM uses a model to generate noise directly for the adversarial examples, rather than clean data. CUDA \cite{sadasivan2023cuda} applies a class-wise noise in the fourier domain for all the samples based on a private key for each class. This tricks the model to not learn any useful information and hence affects generalization. \cite{fang2023towards} propose a method to train a robust surrogate model and use it to generate noise for unlearnable datasets.
\vspace{-2mm}
\subsection{Limitations of existing works}
Important characterstics of unlearnable datasets include robustness to adversarial training, high data quality and low time and compute for creation. 
CUDA addresses an important problem of expensive time compute needed in previous unlearnable dataset creation methods such as REM. However, due to convolving noise over the entire data sample makes CUDA not applicable for practical purposes.

Moreover, existing methods focus dominantly on image datasets. To our best knowledge, we are the first to extend the concept of unlearnability to audio datasets. Methods such as CUDA do not work well in case of temporal data, as applying noise over audio heavily corrupts the data. For example, when an user uploads an audio/music sample on the internet, they would not want to upload a noisy/jittery audio and would like to preserve the quality of the audio/music sample as much as possible. CUDA blurs out the entire audio sample leading to extremely incomprehensible audio samples.

\section{PosCUDA}
\label{sec:poscuda}

We formally motivate the problem of PosCUDA in the context of N-class classification problem. \\
\textbf{Problem}: Given a clean training dataset $DT = \{x_i,y_i\}^n_{i=1}$ and a clean testing dataset $Dte$, a performance objective $P_\theta(DT)$ denoting the performance of a classification model with parameters $\theta$ on dataset $DT$, our goal is to create an unlearnable dataset $\hat{DT}=\{\hat{x_i},y_i\}^n_{i=1}$. The attacker trains a model with parameters $\hat{\theta}$ on $\hat{DT}$. The objective of PosCUDA is to ensure that $P_{\hat{\theta}}(Dte) \ll P_\theta(Dte)$ and $F(\hat{DT},DT) \simeq 0$ is satisfied, where $F(\hat{DT},DT)$ is a similarity score between the two datasets. 
\subsection{PosCUDA: Algorithm}
In this section, we formally explain the PosCUDA algorithm to create unlearnable datasets. PosCUDA applies positional blurs to each data sample. The position of the blur and the filter is determined by a private key for each class. This means that only a small region of the data sample is affected by the blur and majority of the original quality is retained, making it usable for practical purposes. Moreover, there are multiple relationships to the label embedded in the data i.e both the noise and the position of the noise. Hence, adding a very small amount of noise should be sufficient. 
\begin{figure}[t]
    \centering
    \includegraphics[width=8.5cm]{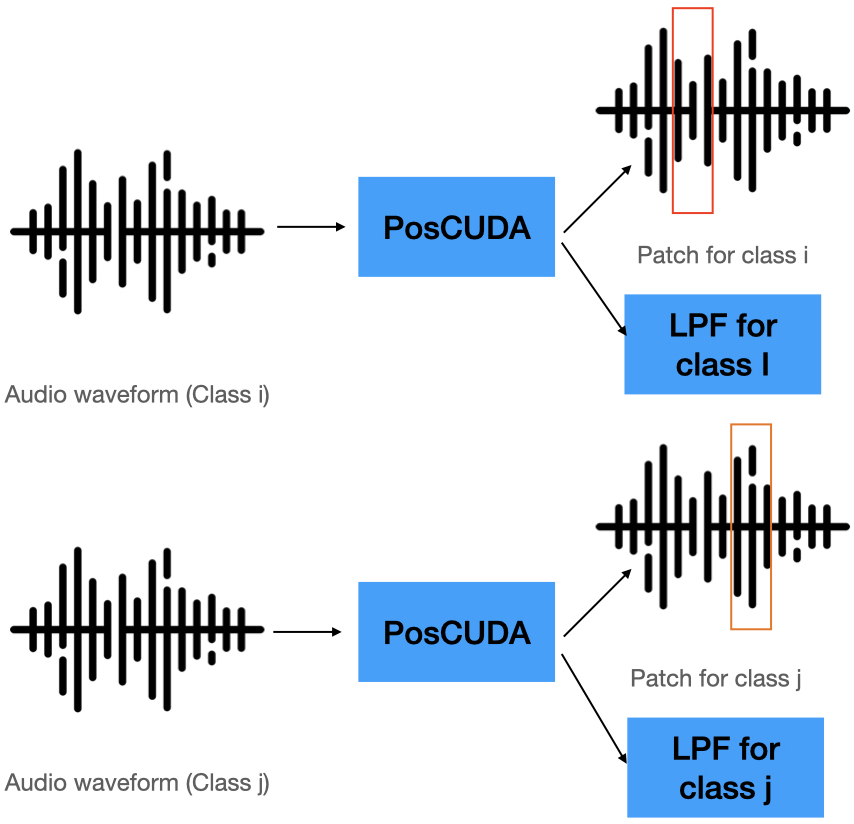}
    \caption{PosCUDA for Audio data: For each of the classes i,j, different patches of audio are passed through a low-pass filter unique to each class. This embeds a small class dependent positional noise in each data sample in the training set. The model learns to map these positional blurs to the labels, failing to generalize in the absence of blurs in the test dataset. }
    \label{fig:enter-label}
\end{figure}
% \begin{figure}[htb]

% \begin{minipage}[b]{1.0\linewidth}
%   \centering
%   \centerline{\includegraphics[width=8.5cm]{images/poscuda.001.jpeg}}
% %  \vspace{2.0cm}
%   \centerline{(a) Result 1}\medskip
% \end{minipage}
% %
% \begin{minipage}[b]{.48\linewidth}
%   \centering
%   \centerline{\includegraphics[width=4.0cm]{images/image3.eps}}
% %  \vspace{1.5cm}
%   \centerline{(b) Results 3}\medskip
% \end{minipage}
% \hfill
% \begin{minipage}[b]{0.48\linewidth}
%   \centering
%   \centerline{\includegraphics[width=4.0cm]{images/image4.eps}}
% %  \vspace{1.5cm}
%   \centerline{(c) Result 4}\medskip
% \end{minipage}
% %
% \caption{Example of placing a figure with experimental results.}
% \label{fig:res}
% %
% \end{figure}
For a given audio data sample belonging to class $i$, we first extract a small patch of the audio of patch size $p$. The location of the patch is specified by a parameter $pos(i)$ denoting the position to extract patch for each class $i$. PosCUDA uses 1-D convolutional filters $f_i$ of size $k$ for each class $i$. These filters are randomly generated for each class based on a private key. We generate these filters from $U(0,b)$ where $b$ is the blur factor. These filters act like low-pass filters for the audio sample. The main assumption is same as that in CUDA, these keys should not be leaked. These filters are applied at the extracted patches for each data sample. We convolve 1D filters to apply these blur filters over the patch. For instance, consider an audio sample $x_i = {a_1,a_2,\dots, a_N}$ belonging to class $i$. Let's assume the location of the patch for class $i$ is $ pos(i) = 100$, patch size $p = 80$, then we first extract the patch of size $1\times80$ starting at position $100$, i.e ${a_{100}, a_{101}, \dots, a_{180}}$ of the sample $x_i$. This means that only a small portion of the audio, in this case, a small patch of $1\times80$ gets blurred, while the remaining of the image retains the original quality. \\
\emph{Understanding PosCUDA:} The main motivation behind PosCUDA is that majority of the machine learning models such as LSTM, Transformers or CNNs learn spatio-temporal relationships in the data. After injecting positional noise, these models are easily able to map the different blurs/noises applied at different positions to the label immediately, failing to generalize to unseen data.

\begin{table*}[h]
    \centering
\begin{tabular}{|c|c|c|c|c|c|c|c|c|}
\hline
\multirow{2}{*}{Dataset} & \multirow{2}{*}{Architecture} & \multirow{2}{*}{Clean}&\multicolumn{2}{c|}{Ours(w/ pos)} & \multicolumn{2}{c|}{Ours} & \multicolumn{2}{c|}{FAD} \\
\cline{4-9}
 & & & b = 0.01 & b = 0.3 & b = 0.01 & b = 0.3 & b = 0.01 & b = 0.3  \\
\hline
 & CNN & 90.52 & 82.30 & 67.33 & \textbf{7.14} & 7.30& &\\
 \multirow{1}{*}{SpeechCommands} & LSTM & 90.74 & 84.63 &  12.95 & 18.32 & \textbf{12.12} & \multirow{1}{*}{$5.61\times10^{-4}$} & \multirow{1}{*}{$5.65\times10^{-4}$}\\ 
 & Transformer & 66.80 & 62.29 & \textbf{13.50} & 60.77 & 32.29 & &\\
\hline
 & CNN & 90.89 & 47.10 & 29.80 & 15.51 & \textbf{10.04} & &\\
 \multirow{1}{*}{FSDD} & LSTM &  96.00 & 45.67 &  42.11 &  32.44& \textbf{22.78} & \multirow{1}{*}{$2.21\times10^{-5}$} & \multirow{1}{*}{$2.25\times10^{-5}$}\\ 
 & Transformer & 91.83 & 72.67 & 53.89 & 62.78 & \textbf{48.11} & &\\
\hline
% etc. ...
\end{tabular}
\caption{Test Accuracy and FAD scores of PosCUDA on SpeechCommands and FSDD datasets. }
\label{table1}
\end{table*}
\section{Experiments}
\label{sec:experiments}
In this section, we first demonstrate the effectiveness of PosCUDA on different audio datasets and various input representations. We also validate PosCUDA empirically on different classification models. We then conduct analysis on the quality of the unlearnable datasets we generate using our method. Finally, we also prove the robustness of PosCUDA under different data augmentation settings.\\
\textbf{Datasets:} We use two main datasets for our experiments:
\begin{itemize}
    \item Speech Commands: SpeechCommands dataset consists of over 100,000 audio samples belonging to 35 classes. Each sample has approximately one second of audio.
    \item Free Spoken Digit Dataset: FSDD datasets consists of 6 speakers and over 3000 samples. Each samples belongs to one class between (0-9). 
\end{itemize}

\textbf{Architecture:} It is important to test the robustness of PosCUDA across multiple architecture choices for audio classification. We also experiment with different input representations such as using raw audio or MFCC co-efficients to validate the robustness of our approach. We run experiments with the following architectures:
\begin{itemize}
    \item M5: We use the M5 CNN architecture from \cite{dai2017very} as one of the networks for audio classification. The M5 architecture consists of a series of 1D convolutions with batch normalization and max pooling. The model takes in raw audio waveform as input. 
    \item LSTM: For the LSTM architecture, we extract the MFCC co-efficients for the audio. We use a one layer LSTM with 128 hidden units for our analysis.
    \item Transformer:  We use a sequence of Encoder blocks from \cite{vaswani2017attention}. For our audio classification task, we use one encoder block with 8 heads and an embedding dimension of 256.
\end{itemize}
\textbf{Evaluation:} We evaluate the effectiveness of PosCUDA primarily on two aspects: 1) Accuracy and 2) Frechet Audio Distance (FAD)\cite{kilgour2019frechet}. The FAD score is similar to Frechet Inception Distance \cite{heusel2017gans} for images. First embeddings of both the clean audio dataset and PosCUDA's polluted dataset is computed using a pretrained model. We use PANN \cite{kong2020panns} to calculate embeddings. FAD is computed by estimating the Fréchet distance between multivariate Gaussians estimated on these embeddings. The lower the FAD score, the more close the polluted dataset is to the clean dataset, with a score of 0 meaning both the datasets are identical. \\
\textbf{Baselines:} To our best knowledge, we are the first to introduce the concept of unlearnability for audio datasets. We introduce another variant of our method without the positional blurs, i.e the low-pass filter operates on the entire data sample. This helps in higlighting the importance of positional blurs.  \\ 
\textbf{Implementational Details:} We run every experiment for 100 epochs. We use filter size $k = 80$  and a patch size of $p=240$ for all our experiments. The blur parameter is a hyperparemeter that can be tuned. We experiment with two settings: a low blur parameter of $b=0.01$ and a high blur parameter of $b=0.3$ for both the datasets. We pollute only the training set for all datasets and keep the test dataset unpolluted. We set the batch size to $128$ and learning rate to $0.01$.
\subsection{Analysis}
Table \ref{table1} shows the performance of PosCUDA compared to other baselines. PosCUDA achieves a low test score on the clean test dataset by a large margin compared to without position variant and the unpolluted dataset. For instance, PosCUDA($b=0.01$) achieves a test accuracy of 7.12\% while the clean test accuracy is 90.52 while not using positional blur achieves 82.30\%. We also see that PosCUDA achieves a low test performance across a wide variety of model architectures such as CNN, LSTM and Transformers, while also being robust to different audio representations such as raw waveforms and MFCCs. An interesting observation is that Transformer architecture performs better compared to other architectures on unlearnable datasets. For example, on the FSDD dataset, the transformer architecture achieves a test accuracy of 48.11\% using PosCUDA($b=0.3$) polluted train set, while CNN and LSTM architectures achieve only 10.04\% and 22.78\% respectively. Transformer architectures also perform slightly better when using the without positional blur variant on the SpeechCommands dataset. Table \ref{table1}  also shows the FAD score to compare the quality of the audio samples. As we can see, PosCUDA retains the maximum quality and achieves a very low FAD score of $2.2x10^-7$ (close to 0). This suggests that PosCUDA provides a very strong unlearnable dataset while maintaining the quality as identical to the original dataset, making it useful for practical purposes.The FAD score also slightly increases when the blur factor is increased to $b=0.3$ but this increase is negligible ($0.01\times10^{-5}$), again validating the importance of positional blurs. Additionally, we observe that the unlearnability efffect increases with increase in the blur parameter. For instance in both the variants, the test accuracy for FSDD decreases as the blur parameter is increased to 0.3. However, PosCUDA achieves a lower test accuracy due to the positional filters. With the same blur parameter of $0.3$, PosCUDA achieves a 10.04\% accuracy for CNN architecture, while non-positional variant achieves 29.80\%.\\
Inorder to validate the hypothesis that PosCUDA maps both the position and the noise to the label, we construct a polluted test dataset for both FSDD and SpeechCommands. The polluted test dataset is constructed in the same manner as the polluted training set, i.e the same positional blurs are applied to each data sample in the test set according to its class. We observe that PosCUDA achieves a 99.8\% in the test dataset under the CNN architecture. This shows that the model has succesfully learnt the mapping between the positional blurs and the corresponding classes. We also try to mix the positional blurs for each class, i.e use the positional blur corresponding to class 1 for class 2 and so on. On such a polluted dataset, the model fails again with a very low test accuracy. This further validates our claims and proves the robustness of PosCUDA. It is not easy for an attacker to randomly add noise at any position and break the unlearnability effect.

% \begin{table}[h]
%     \centering
%     \begin{tabular}{cccccc} \toprule
%     {Dataset} & {Architecture} & {Clean} & {Ours}  & {CUDA} \\ 
%     \midrule
%      \\ 
%      & CNN &90.52 & 12.70 & 81.97 \\
%     \multirow{1}{*}{SpeechCommands} & LSTM & 90.74 & 23.34 & 85.71\\
%     & Transformer & 66.80 & 56.15 & 61.56\\ 
%  \midrule
%      & CNN & 90.89 & 17.19 & 50.11\\
%     \multirow{1}{*}{FSDD} & Bi-LSTM & 96.00 & 70.56 & 63.22\\
%     & Transformer & 91.83 & 55.33 & 62.78\\
%  \\
% \end{tabular}
%     \caption{FAD scores of PosCUDA and CUDA with different blur parameters.}
%     \label{table1}
% \end{table}
\section{Threats}
 Unlearnable datasets face attacks from malicious attackers by attempting to bypass the preotection mechanisms. In this section, we discuss different threat models/scenarios for PosCUDA.\\
 \textbf{Attack 1. Attacker figures out locations of noise in the data.} The attacker can then perturb data during inference at the same locations.\\
 \textbf{Defense}: PosCUDA entangles both the position of the noise and the noise's private key to the label. Since we use a very small blur factor of 0.01, it is not trivial to exactly find the location of the noise. Even if the attacker figures out the location, it is not easy to replicate the original noise used.  We simulated an experiment where the attacker hypothetically figures out the exact locations of the noise for each class and adds random noise to the data, but the model did not perform well on the test dataset. \\
 \textbf{Attack 2. Attacker collects the original unprotected dataset and the protected datasets to train a conditional generative model to recover the position and the noise.} \\
 \textbf{Defense}: It its important for the defender to delete or store the original dataset in a safe place in order to prevent access to the original data. Another strategy can be to regularly change the protection mechanism (different locations and private keys) in case the original dataset is leaked.\\
 \textbf{Attack 3: Attacker applies random blur/noise to the data to try mimic the protection of PosCUDA}\\
 \textbf{Defense}: Unless the attacker obtains the private key of the noise and the locations of the noise for each class label, it is not possible to mimic the protected dataset. We run experiments with random noise added to the test dataset and the model performed poorly on the test dataset. 

 \section{Limitations and Future Work}
 While PosCUDA provides strong robust unlearnable audio datasets, the algorithm has a few limitations. PosCUDA is not ideal for long audio/musical signals, as the attacker can remove the region of noise and use the remaining data for training models. PosCUDA also works only for supervised discriminative models and the effect on unsupervised models is a good future direction to pursue. Another interesting direction would be to extend PosCUDA to generative models. Since most of the generative models such as GANs depend on some form of discriminative models, methods such as PosCUDA can be a good tool to achieve unlearnability. 
\pagebreak

\bibliographystyle{IEEEbib}
\bibliography{refs}

\end{document}